\documentstyle[12pt]{article}
\newcommand{\be}{\begin{equation}}
\newcommand{\ee}{\end{equation}}
\newcommand{\bd}{\begin{displaymath}}
\newcommand{\ed}{\end{displaymath}}
\newcommand{\baa}{\begin{array}{lll}}
\newcommand{\eaa}{\end{array}}
\newcommand{\ba}{\begin{eqnarray}}
\newcommand{\ea}{\end{eqnarray}}

\begin{document}
%\draft
\begin{center}
{\Large\bf A positivity bound for the
 longitudinal gluon  distribution in a nucleon} \\[10mm]

{\bf \large J. Soffer \footnote{E-mail: soffer@cpt.univ-mrs.fr
}}\\

{\it Centre de Physique Th\'eorique - CNRS - Luminy,\\
Case 907 F-13288 Marseille Cedex 9 - France} \\

{ \bf \large and \ \ O. V. Teryaev\footnote{
E-mail: teryaev@thsun1.jinr.dubna.su}} \\

{\it Bogoliubov Laboratory of Theoretical Physics, \\
Joint Institute for Nuclear Research, Dubna, 141980, Russia}
%\date{\today}
%\maketitle
\end{center}
\begin{abstract}
The distribution of longitudinal gluons in a nucleon,
introduced earlier by Gorsky and Ioffe, is
estimated from below by making use of the positivity of density matrix
and the analogue of the Wandzura-Wilczek relation for the light-cone
distributions of polarized gluons in a transversely polarized nucleon.
\end{abstract}
%\pacs{PACS. 12.38.Bx, 13.88.+e}

%\narrowtext
%\section{Introduction}

It is now well known that the spin properties of gluons and quarks are
fairly different. In particular,
there is no analogue of the twist-two
transversity distributions for massless gluons and  their contribution to the
transverse asymmetry starts at the twist-three level.
Also, longitudinal massless gluons do not exist.
However, due to the confinement property, gluons should acquire
an average mass and/or a transverse momentum of the order of the 
inverse of the
hadron size.
As a result, one can have a nonzero longitudinal gluon distribution.
Generally speaking, it is suppressed by the gluon mass squared and 
contributes
at the twist-four  level. However, in the case of Deep Inelastic 
Scattering (DIS),
this is cancelled by the pole in the box diagram.
This effect was studied in details for longitudinal gluons
by Gorski and Ioffe \cite{GI}.
It was shown to be related to the conformal anomaly,
and one may wonder,if it could be observed in other processes.

The gluon mass and/or its intrinsic transverse momentum
should result also in a nonzero transverse gluon distribution.
It was recently studied in \cite{ST97} where the twist-two approximation 
was derived. 
The resulting double transverse spin asymmetries $A_{TT}$, 
for low mass dijet
or low $p_T$ direct photon production,
at RHIC are rather small ($\leq 1\%$) due to a kinematic suppression factor.
%We might have a less severe suppression in the case of single jet
%production in the intermediate $p_T$ region,
%where the quark gluon subprocess dominates.
It seems also promising to study the double
asymmetries
in  open charm or heavy quarkonia leptoproduction
by a longitudinally polarized lepton beam off a transversely polarized
target. The suppression factor that enters in this case is to the first power
of $M/\sqrt{\hat s}$, contrary to the second power for some double
transverse spin asymmetries. The situation seems especially favorable
for the  case of  diffractive
charmonium production,investigated recently \cite{RY}, when the partonic c.m. energy $\sqrt{\hat s}$
is of the order of the charm quark mass.
This would make possible the measurement of the transverse gluon distribution.

The aim of the present paper
is to propose a way to relate these two quantities, generated by off-shell
gluons, namely
the longitudinal gluon distribution and the transverse gluon distribution.
To do this, let us start from the light-cone density matrix of gluon,
namely:

\begin{eqnarray}
\label{dm}
%{1\over{2M}}
\int{{d\lambda \over2\pi}}
e^{i\lambda x}
\langle p,s|A^\rho(0) A^\sigma (\lambda n) |p,s \rangle \sim
{1 \over 2} (e_{1T}^\rho e_{1T}^\sigma+e_{2T}^\rho e_{2T}^\sigma)G(x)
+\nonumber\\
{i \over 2} (e_{1T}^\rho e_{2T}^\sigma-e_{2T}^\rho e_{1T}^\sigma)\Delta G (x)
+
{i \over 2} (e_{1T}^\rho e_L^\sigma-e_L^\rho e_{1T}^\sigma)
\Delta G_T (x) + G_L(x) e_L^\rho e_L^\sigma~,
\end{eqnarray}
where $n$ is the gauge-fixing light-cone vector such that $np=1$,
and we define two transverse polarization vectors,
$e_{1T}$ and $e_{2T}$ . One of them, namely $e_{2T}$ for definiteness,
is chosen to be parallel to the direction of the transverse component of the
polarization, so that's why only the vector $e_{1T}$ enters in the
contribution of $\Delta G_T (x)$. Also, we introduce the longitudinal
polarization vector \cite{GI}

\be
e_L={M\over{\sqrt{(pq)^2-M^2 q^2}}}(q-p{pq \over{M^2}})~,
\ee
for the gluon of momentum $p$ and mass $M$, the latter being taken
equal to that of the hadron.

We denote by
$s_{\mu }$ the covariant
polarization vector  of the proton of momentum
$p$ and mass $M$ and we have $s^2=-1, sp=0$.
Here $G(x)$ and $\Delta G(x)$ are the familiar unpolarized gluon distribution
and gluon helicity distribution, respectively.
The transverse gluon distribution $\Delta G_T(x)$
is the most natural way to measure its transverse polarization,
analogous to the quark structure function $g_T=g_1+g_2$.

The twist-four distribution of longitudinal gluons, as it was mentioned
earlier,is actually contributing
to DIS, at the  twist-two level \cite{GI}. This is because the
relevant box diagram has a pole when the gluon virtuality is
going to zero. Consequently, the mass parameter $M^2$,
which appears in the last
term of (\ref{dm}), is cancelled
being of the same order.

The light-cone distributions $\Delta G(x)$ and $\Delta G_T(x)$ can be
related to each other, if only the twist-two part of $\Delta G_T$
is considered,since as shown in ref.(2),  we have

\begin{eqnarray}
\label{xWW}
\Delta G_T(x)=\int_x^1 {\Delta G (z)\over z}dz~.
\end{eqnarray}

Our present knowledge on $\Delta G(x)$, which is not very precise,
allows a great freedom, so several different parametrizations have been
proposed in the literature \cite{B1,B2,GS,GRSV}.In ref.(2 ), we have shown 
in Figs.1a and 2a,
some possible gluon helicity distributions $x \Delta G(x)$ and in Figs.1b and
2b the corresponding $x \Delta G_T(x)$, obtained by using (\ref{xWW}).
It is worth recalling  from these pictures that, in all cases $x \Delta G(x)$ 
and $x \Delta G_T(x)$ are rather similar in shape and magnitude.

Note that $\Delta G_T(x)$ may be also considered as an analogue
 of transversity for quark,
since for gluons there is no such a difference, caused for quarks by
the two Dirac projections (chiral-even and chiral-odd).
It is proportional to the following gluon-nucleon matrix element

\begin{eqnarray}
\label{hel}
\Delta G_T(x)=
\langle +,1|-,0\rangle
\equiv M_{+-} ~,
\end{eqnarray}
where $(+,-)$ and $(1,0)$ are the nucleon and gluon helicities , respectively.

At the same time we have

\begin{eqnarray}
\label{hel1}
G_L(x)=
\langle +,0|+,0\rangle=
\langle -,0|-,0)\rangle
\equiv M_0 ~.
\end{eqnarray}

To establish the connection with the existing positivity relations,
let us consider
first the forward $\gamma^*p$ elastic scattering  ($\gamma^*$ is
a massive photon), which allows to calculate the DIS structure functions.
It is described in terms of
four helicity amplitudes: $M_0, M_{+-}$ defined above and
\begin{eqnarray}
\label{hel3}
M_+=\langle +,1|+,1\rangle, \, ~ ~  M_-=\langle -,1|-,1\rangle ~.
\end{eqnarray}

There is a well-known condition established long time ago by
Doncel and de Rafael \cite{DDR}, written in the form

\begin{eqnarray}
\label{DDR}
|A_2| \leq \sqrt{R}~,
\end{eqnarray}
where  $A_2$ is the usual transverse asymmetry and
$R=\sigma_L/\sigma_T$ is the standard ratio in DIS. It reflects a non-trivial
positivity condition one has on the photon-nucleon helicity amplitudes, which read using the above notations

\begin{eqnarray}
\label{pdis}
(M_{+-})^2 \leq 1/2(M_+ + M_-) M_0 ~.
\end{eqnarray}

One can apply this result to the similar case of
gluon-nucleon scattering, adding to the definitions (4,5,6),
the following ones

\begin{eqnarray}
\label{helg}
G(x)=M_+ + M_- , ~ ~ \Delta G(x)=M_+ - M_- ~.
\end{eqnarray}

As a result, the positivity relation (8) leads to
\begin{eqnarray}
\label{ineq}
|\Delta G_T(x)| \leq \sqrt{1/2G(x)G_L(x)}~.
\end{eqnarray}

It is most instructive to use this relation to estimate
$G_L$ from below
.

\begin{eqnarray}
\label{bound}
G_L(x) \geq 2[\Delta G_T(x)]^2/G(x)  = 2\lambda(x)G(x) ~,
\end{eqnarray}
where $\lambda(x)= [\Delta G_T(x)/G(x)]^2$.

Note that given the data on $R$ in DIS, one obtains from (7) an
{\it upper bound} on $|A_2|$, which is satisfied
by polarized DIS data \cite{AB}, and is far from saturation.
However (\ref{bound}) provides a {\it lower bound} on $G_L(x)$
since $G(x)$ is known from unpolarized DIS  or direct photon
production and $\Delta G_T(x)$ can be evaluated \cite{ST97},
in the twist-two approximation if one  uses eq.(3).
One obtains $\lambda(x) \simeq 0.01$ for $x \simeq 0.1$ or so,
and our lower bound gives $G_L(x) \geq 0.3$ or so.
For lower $x$ values, due to the rapid rise of $G(x)$,
$\lambda(x)$ is much smaller and, for example, for $x \simeq 10^{-3}$,
we find $G_L(x) \geq 10$ or so. At the same time, for very large
$x \rightarrow 1$, since $ \Delta G_T(x) $ is similar to $\Delta G(x)(1-x)$,
$\lambda$ is close to zero, and $G_L(x)/G(x) \to 0$.

Such a relation is of special interest, since it relates, at least formally,  different twist structures. This is by no means surprising,
because, in the case of polarized DIS, if we would have known
$A_2$ before $R=\sigma_L/\sigma_T$,
we could have also estimated the latter from below.
Note that physically the existence of such a relation
is due to the fact, that transverse and longitudinal gluon distributions
are generated by the same source, the gluon mass.

We are indebted to B.L. Ioffe, A. Khodjamirian,
A. Sch\"afer and D. Sivers for
valuable comments and interest in the work.
This investigation was supported in part by INTAS Grant 93-1180 and by the
Russian Foundation for Fundamental Investigations under Grant 96-02-17631.

%newpage
%\begin{center}
%\large{Figure Captions}
%\end{center}

\end{document}